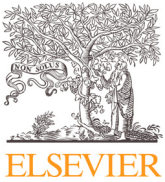
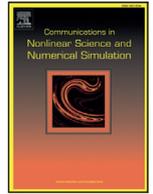

Research paper

# Stochastic 0-dimensional Biogeochemical Flux Model: Effect of temperature fluctuations on the dynamics of the biogeochemical properties in a marine ecosystem

Paolo Lazzari [a], Roberto Grimaudo [b,*], Cosimo Solidoro [a], Davide Valenti [b]

[a] *National Institute of Oceanography and Applied Geophysics - OGS, Italy*
[b] *Dipartimento di Fisica e Chimica "Emilio Segrè", Group of Interdisciplinary Theoretical Physics, Università degli Studi di Palermo, Palermo, Italy*

**A B S T R A C T**

We present a new stochastic model, based on a 0-dimensional version of the well known biogeochemical flux model (BFM), which allows to take into account the temperature random fluctuations present in natural systems and therefore to describe more realistically the dynamics of real marine ecosystems. The study presents a detailed analysis of the effects of randomly varying temperature on the lower trophic levels of the food web and ocean biogeochemical processes. More in detail, the temperature is described as a stochastic process driven by an additive self-correlated Gaussian noise. Varying both correlation time and intensity of the noise source, the predominance of different plankton populations is observed, with regimes shifted towards the coexistence or the exclusion of some populations. Finally a Fourier analysis carried out on the time series of the plankton populations shows how the ecosystem responds to the seasonal driving for different values of the noise intensity.

## 1. Introduction

Marine ecosystem models describing the lower trophic levels of the food web and ocean biogeochemical processes [1–3] are increasingly used in large scale monitoring and assessing infrastructures (e.g. Copernicus Marine Environmental Monitoring Service - CMEMS) to analyze and predict the state of seas and oceans [4]. They are a basic reference for important political decisions regarding essential issues of modern human society, such as climate change, ecosystem health and food provision, unfolding over different space-time scales [5–8].

Global ocean biogeochemical models, for example, give the opportunity of deeply understanding the mechanisms at the basis of carbon cycle, paving the way to a more informed and therefore effective decision management about action to be taken about climate and its effects on marine ecosystems. The role played by oceans in the carbon cycle and in its adjustment is indeed fundamental due to their capability of absorbing atmospheric $CO_2$ and possibly mitigating global warming.

* Corresponding author.
*E-mail addresses:* plazzari@inogs.it (P. Lazzari), roberto.grimaudo01@unipa.it (R. Grimaudo), csolidoro@inogs.it (C. Solidoro), davide.valenti@unipa.it (D. Valenti).

Since the very simple initial attempt [9,10] the interest towards the description of the carbon cycle and other critical aspects of the biogeochemical state of the oceans through complex models has rapidly increased [1,11–14].

A wide class of complex models, used to investigate and predict the physical, biogeochemical, and ecological properties of marine ecosystems across a wide range of scales, has indeed been devised over the years. In this class it is possible to distinguish the multi-nutrient, multi-plankton biogeochemical models, which can now be considered the state-of-the-art, including the BFM [15], ERSEM [16], PISCES [17], ERGOM [18], and DARWIN [19].

The context of application of BFM includes short-term forecasting [20,21], ocean acidification [22], climate change [23,24], process studies [25–27], biogeochemical cycling [28], and carbon sequestration [29]. It is worth noting that all the models previously cited share the same basic characteristic: they are deterministic models. Nevertheless, real systems are unavoidably coupled with a surrounding environment which is characterized by an intrinsic stochasticity. This noisy interaction can be taken into account by considering the presence of noise source in the dynamics of the systems analyzed.

In the last decades researchers have widely and deeply investigated the effects of random fluctuations on the dynamics of several natural systems [30–32]. The fields of interest range from population dynamics [33–35], infective desease and epidemics [36,37], to bioinformatics [38,39], neuroscience and biological evolution [40,41]. Relevant dynamical effects induced by the presence of stochastic processes have been also brought to light in natural ecosystems [42], in the bacterial growth in food products [43], and in the inception and development of diseases due to genetic mutations [44,45].

All these examples indicate that noise should not be considered as only a source of disorder, but as a necessary and fundamental ingredient to correctly describe the dynamics of open systems and to unveil intriguing and counterintuitive effects.

Remarkable examples of such phenomena in living systems are stochastic resonance [46–49], noise enhanced stability [50,51], and noise delayed extinction [34,52]. The underlying reason of these unpredictable dynamics can be ascribed to the fact that ecological systems are characterised by the simultaneous presence of nonlinear interactions and random fluctuations, which are responsible for complex behaviours. In particular, marine ecosystems are an interesting and, from a socio-economic and ecological point of view, important example of natural complex systems, characterized by nonlinear interactions [53]. Moreover, their dynamics is highly affected by both deterministic forcings (daily and seasonal cycles) [26,54,55] and random fluctuations of environmental variables [30–32,34,37,41,42,52,56–58] such as the temperature [59,60].

The critical role played by environmental parameters in determining the steady-state of an aquatic ecosystem is clearly showed in phytoplankton population dynamics. In this case, for example, physical variables, such as temperature, can modify the spatio-temporal behaviour of the net growth rate of the phytoplankton biomass production mechanism [61]. Moreover, changes of limiting factors such as the light intensity and nutrient concentration cause the phytoplankton system to undergo a passage from a stability condition to another and vice-versa [62,63].

As a consequence, to correctly and exhaustively model the dynamics of a marine ecosystem one has to consider the random perturbations coming from the environment. Devising stochastic models to describe the dynamics of biogeochemical properties is an innovative use of tools already applied to other fields, which allows for a significant advance in the theoretical study of marine ecosystems. Thus, in view of a combined model/experiment approach, stochastic models represent a powerful tool suitable to capture and predict, for the first time, the dynamics of biogeochemical properties of real marine ecosystems.

In light of these considerations and according to previous works [56,64–68], the model used in this work takes into account the effects of random fluctuations by considering the temperature as a stochastic process. We have devised a zero-dimensional stochastic biogeochimical flux model (SBFM), starting from the well known deterministic biogeochemical flux model (BFM)[69]. The latter is used in a wider context to simulate the biogeochemical dynamics as driven by circulation, seawater properties, photosynthetically active radiation, and biogeochemical interactions. The BFM simulates a planktonic food web of four phytoplankton populations, one heterotrophic bacterial group, two microzooplankton groups, and two mesozooplankton groups. It accounts for biogeochemical cycles of carbon, phosphorus, nitrogen, and silicate. The evolution of dissolved and particulate organic matter is also included, as well as the dynamics of the microbial loop. As claimed above, the SBFM includes additive noise sources to simulate the stochastic variability of physical variables such as the temperature. In particular, the temperature dynamics has been modeled as that of a Brownian particle subject to an additive self-correlated Gaussian noise.

The study shows that significant effects on the low-trophic-chain dynamics are induced by the presence of random fluctuations, which strongly influence the relative population concentrations present in the ecosystem.

The paper is organized as follows: in the Modelling section we present both the deterministic and the stochastic BFM. In the results section we illustrate and discuss the outcomes of numerical simulations designed to estimate the effects on the trophic web. The last section is devoted to final remarks and conclusions.

## 2. Modelling

The BFM was specifically designed to explore biogeochemical marine ecosystems. In general it is used to describe the dynamics of chemical species and biological populations, i.e. plankton functional types, through deterministic coupled Fokker-Planck equations. The BFM allows also to describe the dynamics of physical variables, with all ecosystem parameters being set according to experimental and/or theoretical data reported in literature.

In this work the temperature, an environmental variable of paramount importance, due to its regulatory role in several chemical, physical and biological mechanisms, is modeled as a stochastic process. The noisy dynamics of temperature is implemented by including its random fluctuations. The original BFM becomes therefore a Stochastic Biogeochemical Flux Model (SBFM).

Other models have been devised to study stochastic dynamics and noise induced effects in interacting population systems [34,35,40]. These approaches however catch partially the complexity of marine ecosystems, whose dynamics is the result of the interplay between a strongly nonlinear biogeochemical dynamics within a multi-level trophic chain and the random fluctuations of environmental variables such as the temperature.

The SBFM therefore is able to account for specific interactions and mechanisms typical of a complex trophic network as well as the noisy dynamics induced in marine ecosystems by a continuous exchange with the surrounding environment. The SBFM indeed allows to model the dynamics of a real marine ecosystem, including both deterministic (typically daily and seasonal) drivings and randomly fluctuating perturbations coming from the environment. The complexity of real marine ecosystems requests an adequate modeling, resulting in a highly complex mathematical structure such as that of the SBFM, which includes both the characteristics of the BFM and the effects of environmental random fluctuations. A detailed description of both BFM and SBFM is given in the following sections.

## 2.1. The Biogeochemical Flux Model

The BFM is a deterministic biomass-based model developed to reproduce the major biogeochemical processes occurring in the pelagic marine ecosystems. The BFM simulates the cycles of nitrogen, phosphorus, silica, carbon, and oxygen in water due to plankton activity. The BFM plankton functional types (PFTs) are subdivided in primary producers (phytoplankton), predators (zooplankton), and decomposers (bacteria), classified on the basis of prescribed functional traits. These broad functional classifications are further partitioned into functional subgroups which contribute to define the planktonic food web. Producers are divided in four classes: diatoms (P1), flagellates (P2), picophytoplankton (P3) and dinoflagellates (P4). Microzooplankton and mesoozooplankton compose the predator group. Microzooplankton class is in turn partitioned in heterotrophic nanoflagellates (Z6) and microzooplankton (Z5). Mesozooplankton is divided in omnivorous mesozooplankton (Z4) and carnivorous mesozooplankton (Z3). Bacteria (B1) are responsible for the recycling of organic compounds in inorganic constituents such as nitrates, phosphates and silicates. The trophic web is schematised in Fig. 1. Hence, the BFM provides a much more realistic description of the microbial food-web trophic interactions than the remarkably simpler Lotka-Volterra type models [70–72] with a food chain composed by two species (one prey and one predator).

All the primary producers (P1, P2, P3, P4) assimilate dissolved inorganic carbon, inorganic nutrients (nitrogen, phosphorus and silicate) and require light to thrive. Large part of the biogeochemical activity in marine ecosystems is driven by these organisms that sustain the trophic web and are responsible for the biogeochemical cycling. Diatoms (P1) have an equivalent spherical diameter (ESD) ranging from 20 to 200 μm, and are unicellular eukaryotes enclosed by a silica protec-tive shell (frustule); diatoms, therefore, need silicates to grow. They are eaten by micro- and mesozooplankton. Autotrophic nanoflagellates (P2), ESD = 2 − 20 μm, motile unicellular eukaryotes, including smaller dinoflagellates and other autotrophic microplanktonic flagellates, are predated by heterotrophic nanoflagellates, micro- and mesozooplankton. Picophytoplankton (P3), ESD = 0.2 − 2 μm, are autotrophic bacteria, they have higher affinity versus ammonia, and are mostly preyed by heterotrophic nanoflagellates. Large, slow-growing phytoplankton (P4), ESD > 100 μm, includes a wide group of phytoplankton species, among which larger species belonging to the previous groups (for instance large flagellates) but also species that during some periods of the year develop a form of biochemical defense to predator attack. This group generally has low growth rates and small food matrix values with respect to micro- and mesozooplankton groups. Bacteria (B1), which include a wide group of aerobic and anaerobic heterotrophic bacteria, exploit, as a source of carbon, organic matter produced by themselves and other organisms. They are able to use organic nutrients or compete with primary producers for the uptake of inorganic matter. They ultimately depend on primary producers, because they are not able to perform photosynthesis and produce organic carbon. Mesozooplankton (Z3, Z4) are consumers with body length between 200 μm and 3 to 4 cm, microzooplankton (Z5) has an ESD from 20 to 200 μm and nanoflagellates are the smallest predators (ESD from 2 to 20 μm). The state of each plankton functional group is defined by a set of components, for example the physiological condition of diatoms is expressed in terms of intracellular concentration of carbon, nitrogen, phosphorus, silicon, chlorophyll. The leading element of the BFM model is carbon that is the fundamental "currency" of life. Nonetheless, the health of the cell is defined by the intracellular quota of elements and stoichiometric balacing versus carbon [73]. Overall the BFM consists of a system of 54 nonlinear ordinary differential equations (ODEs), with a corresponding 54-dimensional state vector ($V_{bfm}$). The derivatives of the generic primary producer scalar carbon component (e.g. carbon in diatom, $P1_c$) and of the corresponding nutrient component (e.g. nitrogen in diatom, $P1_n$) are defined as:

$$\frac{\partial P1_c}{\partial t} = f_{gpp}(V_{bfm}, T, I) - f_{rsp}(V_{bfm}, T, I) - f_{exc}(V_{bfm}, T) - f_{prd}(V_{bfm}, T), \tag{1}$$

$$\frac{\partial P1_n}{\partial t} = f_{upt}(V_{bfm}, T) - f_{rel}(V_{bfm}, T) - f_{prd}(V_{bfm}, T). \tag{2}$$

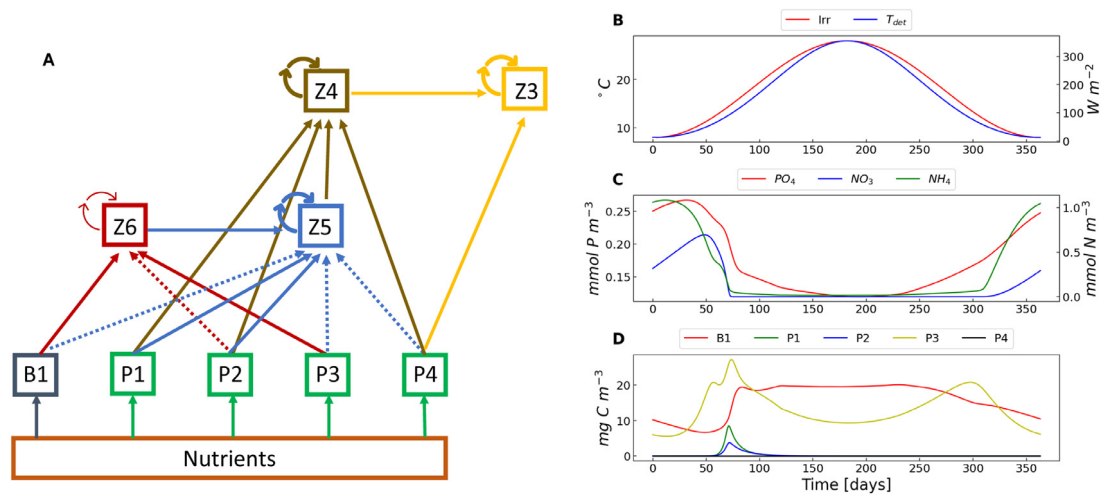

**Fig. 1.** A) Trophic web interactions in the BFM. Nine plankton populations are considered: carnivorous mesozooplankton (*Z3*), omnivorous mesozooplankton (*Z4*), microzooplankton (*Z5*), heterotrophic nanoflagellates (*Z6*), bacteria (*B1*), diatoms (*P1*), nanoflagellates (*P2*), picophytoplankton (*P3*), and dinoflagellates (*P4*). An arrow directed from one box to another indicates a predation flux. Solid arrows denote a higher preference for a specific prey while dashed ones indicate a lower preference. A looping arrow on the box itself denotes cannibalism. Deterministic time behaviour during one year of simulation of: B) temperature ($T_{det}$, °C) and irradiance ($Irr$, $W\ m^{-2}$); C) inorganic nutrients phosphates ($PO_4$, $mmol\ P\ m^{-3}$), nitrates ($NO_3$, $mmol\ N\ m^{-3}$), and ammonia ($NH_4$, $mmol\ N\ m^{-3}$); D) bacteria, diatoms, nanoflagellates, picophytoplankton, and dinoflagellates, all expressed in $mg\ C\ m^{-3}$.

The $f$-s are continuous functions representing various biogeochemical fluxes associated with physiological processes. T and I are environmental temperature and downward solar irradiance, respectively. The *gpp* is the gross primary production (expressed in $mgCm^{-3}day^{-1}$), it is the carbon entry-point in the ecosystem, and is basically related to photosynthesis. Respiration (*rsp*) is related to the release of carbon (production of $CO_2$) and combines basal and activity terms. Excretion (*exc*) processes are related to the cell metabolic activity and to the balancing of the internal quota of carbon versus other elements. In particular, in case of internal nutrient shortage, the organic carbon produced is released as dissolved organic carbon (DOC). The $f$-s are factorised in a number of regulating functions

$$f_{gpp}(V_{bfm}, T, I) = r_{max} f_T(T) f_I(I) f_{nut}(V_{bfm}) P1_c, \tag{3}$$

where $r_{max}$ is the maximum growth rate and it is phytoplankton specific, and $f_I(I)$ is a light harvesting factor dropping to zero in absence of light. The temperature regulation factor $f_T(T)$ is based on an exponential formulation [74]

$$f_T(T) = Q^{\frac{T-T_{ref}}{T_{ref}}}, \tag{4}$$

where Q is 2.00 (non-dimensional) for all the organisms with the exception of heterotrophic bacteria (B1), which have Q equal to 2.95 [75–78]. The reference temperature ($T_{ref}$) is set to 10 °C. The same temperature-regulating factor ($f_T(T)$) acts on the other physiological processes. As a consequence, higher/lower temperatures increase/decrease the overall metabolic rates, accelerating/slowing down the system biogeochemical cycling.

According to Ref. [76], we wish to point out that in reality the temperature dependence of phytoplankton growth rates can be much more complicated as, for instance, the functional relationship is usually nonmonotonic with a species-specific optimal temperature. Nonetheless, as a first step of this new (stochastic) approach to the modeling of real ecosystems, we have chosen to investigate the "zero-order" (monotonic) dependence of $f_{gpp}(V_{bfm}, T, I)$ on the temperature.

The equations for zooplankton are similar to those for phytoplankton, with the gross primary production term replaced by the grazing ($f_{gra}$). In the case of carnivorous mesozooplankton we obtain

$$\frac{\partial Z3_c}{\partial t} = f_{gra}(V_{bfm}, T, I) - f_{rsp}(V_{bfm}, T, I) - f_{rel}(V_{bfm}, T) - f_{prd}(V_{bfm}, T). \tag{5}$$

The total amount of food available to zooplankton is computed summing over the possible preys (see Fig. 1) weighted according to the predator's food preferences. Predation follows a type 2 form [79]

$$f_{gra}(V_{bfm}, T, I) = f_T(T) r_{max} \frac{F_c}{F_c + h_z} Z3_c, \tag{6}$$

$$F_c = \sum_{prey} \delta_x e_{prey} X_i, \tag{7}$$

where $h_z$ is inversely related to the searching volume of the organism, $X_i$ is the carbon content in the preys, $\delta_i$ is the preference for a specific prey, and $e_{prey}$ is the capture efficiency.

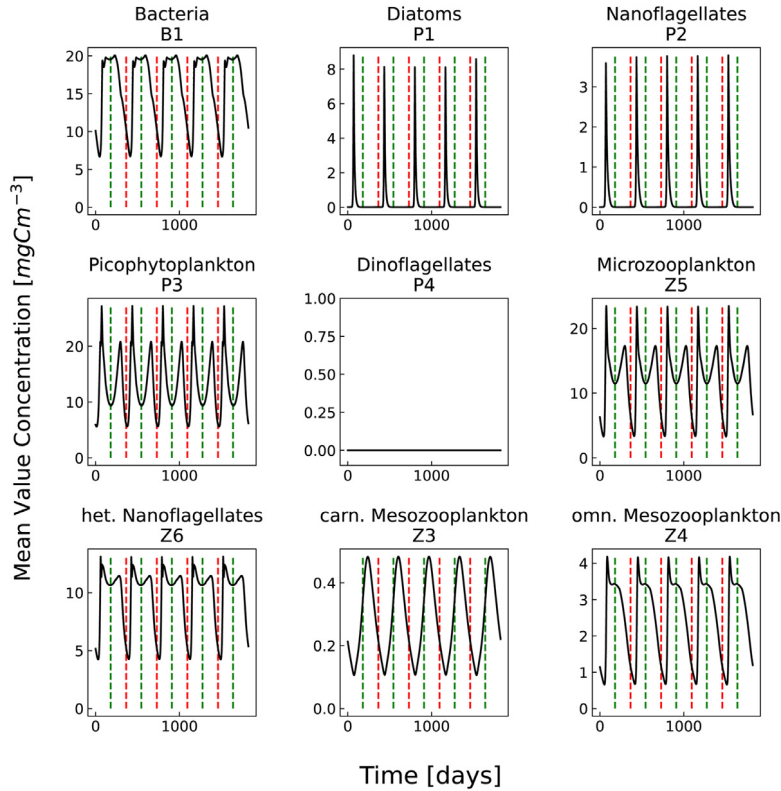

**Fig. 2.** Results from the deterministic 0-D BFM. The different panels show the time evolution of carbon intracellular content ($mgC/m^3$) of the nine plankton populations present in the ecosystem. The time series describe the last 5-years behaviour of a 10-year numerical integration of the model equations, after the ecosystem reached an oscillating steady state. Vertical lines correspond to January $1^{st}$ (red dashed lines) and July $1^{st}$ of each year (green dashed lines).

The full list of equations and processes included in the BFM can be found in [54] and the BFM code manual in [15].

The 0-D model configuration assumes that the organisms are confined in a perfectly mixed water mass. Temperature ($T(t)$) includes a seasonally varying deterministic components ($T_{det}(t)$), with one-year period, minimum temperature in winter (8 °C) and maximum temperature in summer (28 °C). Similarly, downward solar irradiance oscillates with values, lower during winter (20 $Wm^{-2}$), higher during summer (300 $Wm^{-2}$), with a superimposed day-night light cycle.

The deterministic configuration of the BFM model has been used in a number of studies and applications [20–29], to cite a few.

The reference deterministic model presents a cyclic behaviour in all the interacting plankton functional types, Fig. 2. This oscillation is at the steady state related to the seasonal cycle of solar irradiance and environmental temperature. In winter the system is limited by light. When light is no longer limiting, picophytoplankton wins the competition for nutrients (in particular nitrates), while dominating among the primary producers (see panels B, C and D in Fig. 1). Conversely, after the cold period (winter) diatoms and nanoflagellates show very low concentrations. Moreover, due to their growth rate lower than those of the other PFTs, dinoflagellates exponentially decays till reaching zero concentration. The microbial loop compartment consisting of bacteria and picophytoplankton shows sustained biomass during all the year, modulated by the predation in summer, according to the trophic web shown in Fig. 1. At the top of food chain, carnivorous mesozooplankton (Z3) shows a sinusoidal time behaviour (see Fig. 2). Without seasonal forcing-induced variability, the system reaches an equilibrium without intrinsic oscillations (image not shown). Concerning this point, we wish to recall that the ecosystem analyzed consists of three different throphic levels, including overall nine populations, and numerous interactions. As a consequence, even if we can not exclude that, for a certain parameter range, self-sustained oscillations can appear, anyway the high level of complexity of our model, due to both the nonlinearity of equations and the numerous interactions which charachterize the BFM, suggests that the appearance of self-sustained oscillations, typical of a Lotka-Volterra prey-predator system, could be inhibited for any choice of the parameter values.

## 2.2. The Stochastic Biogeochemical Flux Model

Here we modify the initial (deterministic) BFM by considering the temperature as a stochastic process, i.e. a virtual Brownian particle "moving" under the influence of a self-correlated Gaussian noise, modeled through a term of additive noise, which accounts for random fluctuations coming from the environment and always present in a real ecosystem. Mathemati-cally the temperature is described by the following Langevin equation

$$T(t) = T_{det}(t) + F_{TEMP}(t), \tag{8}$$

$$\frac{dF_{TEMP}(t)}{dt} = -\frac{F_{TEMP}}{\tau} + \xi_T(t), \quad F_{TEMP}(0) = 0, \tag{9}$$

where $\xi_T$ is a white Gaussian noise with mean value and correlation function given by

$$<\xi_T(t)> = 0; \quad <\xi_T(t)\xi_T(t')> = D\delta(t-t'), \tag{10}$$

with $D$ being the noise intensity and $F_{TEMP}$ the temperature fluctuation. We recall that the time series of populations in real ecosystems are noisy and specifically they can be "red", i.e. they are dominated by low frequencies (long term variations), or "white", i.e. they present no dominating frequency [80–82]. In particular, "terrestrial populations exist in a white noise atmosphere, whereas marine populations are embedded in a red noise environment" (see Ref. [83,84]). Moreover, we recall that, as one can expect, when red or blue environmental noise sources are present, the time series of populations show, respectively, more red or blue spectra than those subjected to white noise [85]. Many ecologists recognized that environmental noises can be better modeled by auto-correlated noise sources [35]. These experimental and theoretical studies suggest that red noise sources allow to better describe the dynamics of real marine ecosystems.

The fluctuation of temperature described by Eq. (9) corresponds to a damped Brownian motion, the so called Ornstein-Uhlenbeck process. The damping strength is parametrised by the temporal scale $\tau$. $D$ is the amplitude of the noise fluctuation expressed in $°C^2/s$. Mean value and variance of the Ornstein-Uhlenbeck process can be determined analytically [86]:

$$<F_{TEMP}(t)> = <F_{TEMP}(0)> e^{-\frac{t}{\tau}}, \tag{11}$$

$$\sigma_T^2 = \text{var}\{F_{TEMP}(t)\} = \left\{\text{var}\{F_{TEMP}(0)\} - \frac{D\tau}{2}\right\} e^{-2\frac{t}{\tau}} + \frac{D\tau}{2}, \tag{12}$$

where $\sigma_T$ is the amplitude of the temperature fluctuations. For sufficiently long times ($t > \tau$) the average of the fluctuation is zero and the variance stabilises to $\frac{D\tau}{2}$. The stationary correlation function [86]

$$<F_{TEMP}(t)F_{TEMP}(t')>_s = \frac{D\tau}{2} e^{-\frac{|t-t'|}{\tau}}, \tag{13}$$

can be used to compute the spectrum (S) according to the Wiener-Khinchin theorem [86]

$$S(\nu, \tau, D) = \frac{D\tau}{2} \int_{-\infty}^{+\infty} e^{-\frac{|s|}{\tau}} e^{-i2\pi\nu s} ds = D\tau \frac{\tau}{1 + (2\pi\nu\tau)^2}, \tag{14}$$

where $s = t - t'$ is the lag time and $\nu$ the frequency. We note that $\partial_\tau S$ and $\partial_D S$ vanish only for $\tau = 0$ *days*, which is a degenerate condition. On the contrary, along the isoclines of the variance of temperature fluctuations, $\text{var}\{F_{TEMP}(t)\} = c = \frac{D\tau}{2}$, there is a local maximum of power density at frequency $\nu_{peak} = \frac{1}{2\pi\tau}$ [35].

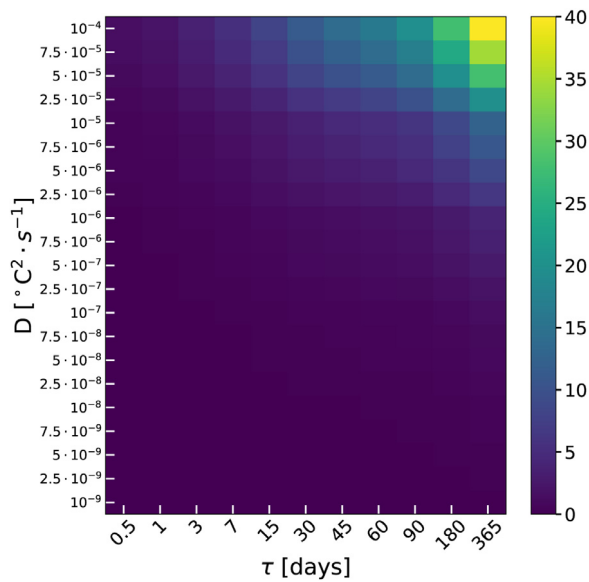

**Fig. 3.** Standard deviation ($\sigma_T$, °C) of temperature according to Eq. (15) for each $D - \tau$ pair considered.

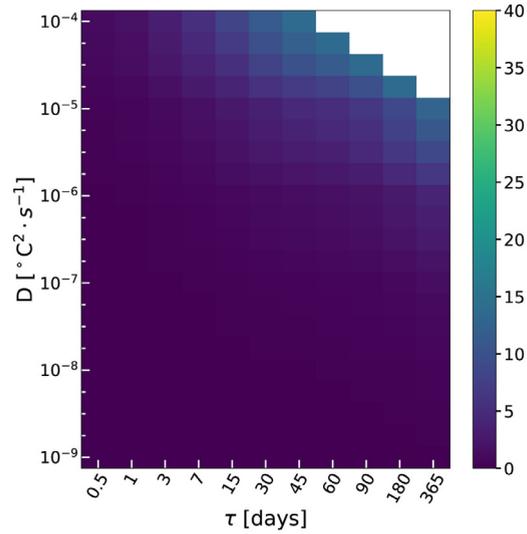

**Fig. 4.** Standard deviation of temperature resulting from numerical integration of model equations for each $D - \tau$ pair considered. Blank pixels are related to the $D - \tau$ region not considered in our analysis because of a high percentage ($> 10\%$) of failed numerical realizations.

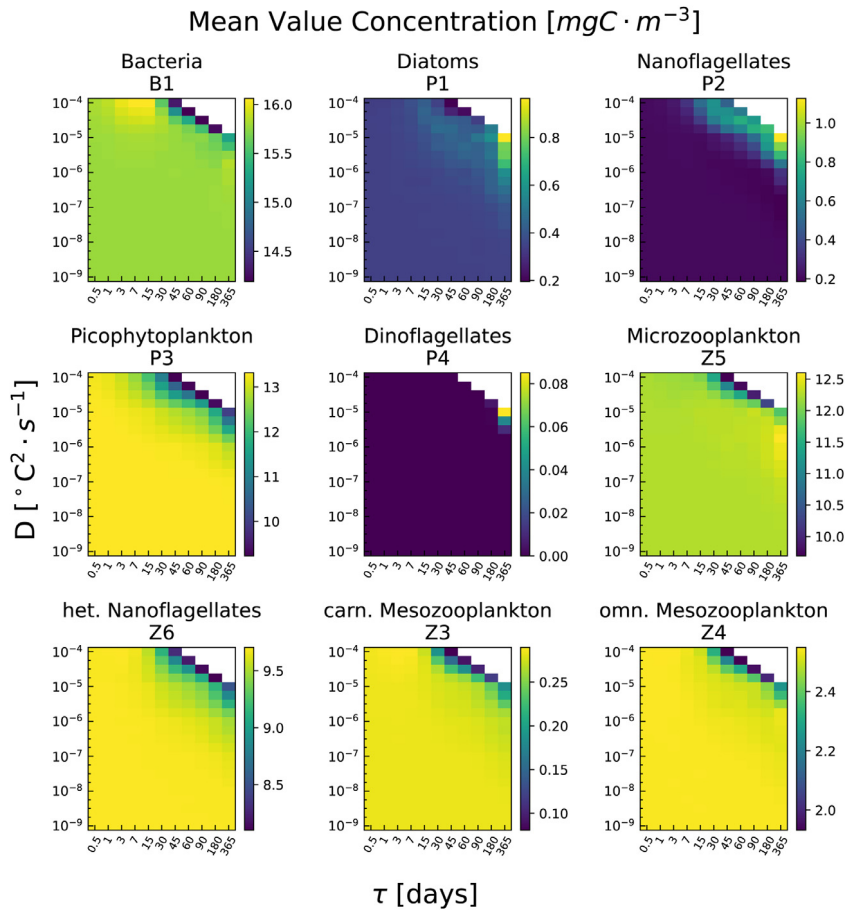

**Fig. 5.** Results from the stochastic 0-D BFM. Each panel shows the mean value (the average is taken over time and over 1000 realizations) of carbon intracellular content ($mgC/m^3$) of BFM plankton functional type with respect to stochastic scenarios corresponding to 231 different $D - \tau$ pairs.

The introduction of the damping term is extremely important. In fact, perturbing the system by modeling the temperature as a Brownian particle, i.e. subject to white Gaussian noise, results in a divergence of the temperature variance in time. Selecting different values of $D$ and $\tau$ allows to estimate how a prescribed variance of temperature affects biogeochemical dynamics.

## 3. Results and discussion

### 3.1. Numerical simulations

The model equations were integrated within a time-window of ten years. In the multi-panel figures shown in the following sections, only the time behaviour during the last five years has been plotted in order to focus the analysis on the oscillating steady state, which characterizes the long-time system behaviour, neglecting the transient dynamics.

Because of the stochastic nature of the model, the concentrations for the different plankton species investigated have been obtained by solving numerically the model equations within the Ito scheme (see Ref. [86], pages 83-86, where the Ito scheme is explained in detail and compared with the alternative definition of stochastic integration introduced by Stratonovich), and averaging over 1000 realizations.

As this point is concerned, it is worth underlining that the effects of noise, i.e. of random environmental fluctuations, are not ruled out by taking the average over the different realizations. This aspect, peculiar of natural systems due to their nonlinearity, is reproduced by the SBFM.

As previously said, the stochastic dynamics is characterized by the self correlation time, $\tau$, and the noise intensity, $D$, which determine the variance of the temperature, $\sigma_T^2(t)$, over the time, according to the analytical expression [86]

$$\sigma_T^2(t) = (1 - \exp\{-2t/\tau\}) \frac{D\,\tau}{2}, \tag{15}$$

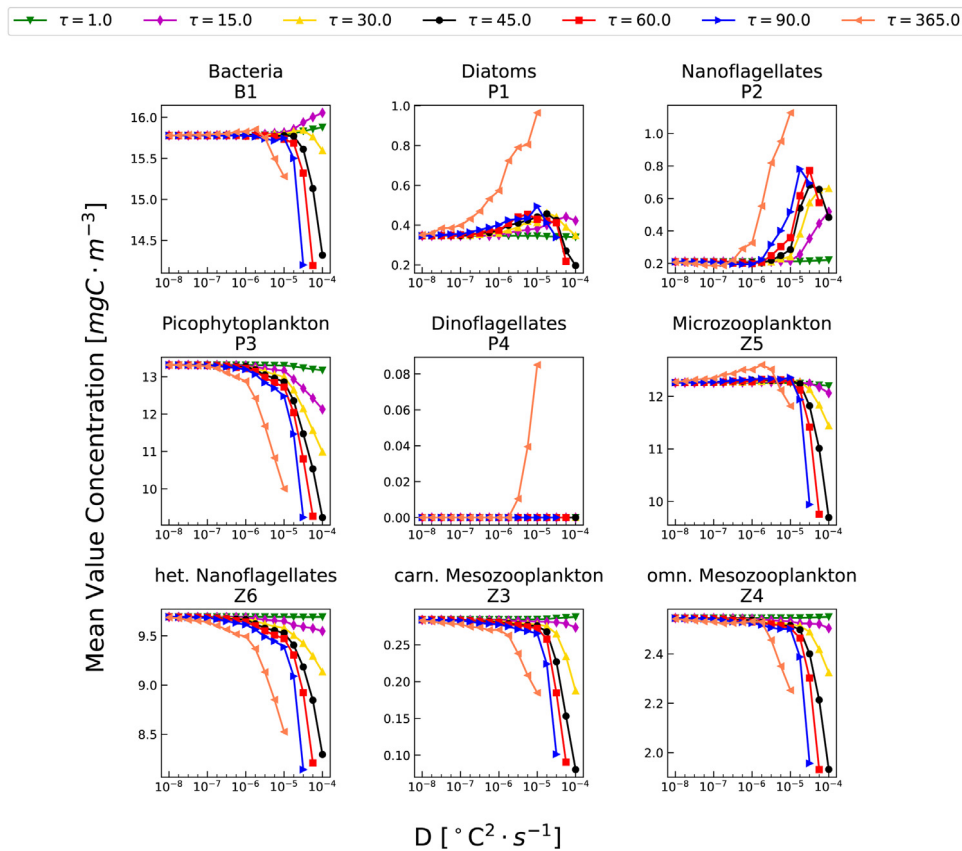

**Fig. 6.** Results from the stochastic 0-D BFM. Each panel shows the mean value (the average is taken over time and over 1000 realizations) of carbon intracellular content ($mgC/m^3$) of BFM plankton functional type with respect to the noise intensity parameter $D$ and for seven different values of $\tau$, namely: $\tau = 1$ *day* (green line, down-triangle markers), $\tau = 15$ *days* (purple line, diamond markers), $\tau = 30$ *days* (yellow line, up-triangle markers), $\tau = 45$ *days* (black line, circle markers), $\tau = 60$ *days* (red line, square markers), $\tau = 90$ *days* (blue line, right-triangle markers), $\tau = 365$ *days* (orange line, left-triangle markers).

whose values at $t = t_{max} = 10$ years are plotted in Fig. 3 for 11 different values of correlation time and 21 different values of noise intensity.

For high values of both $\tau$ and $D$ (upper right corner) the $\sigma_T(t)$ reaches values as high as 35 °C.

Anyway, this parameter region corresponds to cases for which more than 10% of numerical realizations of Eq. (9) fails, this numerical failure being related to a predicted negative concentration (so called false negative). This occurs when the values of $F_{TEMP}$ become too large, i.e. for $D > 10^{-5}$ °$C^2 s^{-1}$ and $\tau > 32$ *days*. In principle, the numerical failure could be solved by reducing the integration time step. Anyway we did not repeat numerical integration by using a smaller time step, which would result in an unnecessary increase of the simulation time. Indeed, for $D > 10^{-5}$ °$C^2 s^{-1}$ and $\tau > 32$ *days* the asymptotic amplitude, $\sigma_T$, of the temperature fluctuations reaches values too large for real ecosystems. Thus, we disregard these points of the $D-\tau$ parameter space, retaining only numerical solutions corresponding to fluctuating temperature with $\sigma_T(t_{max}) \leq 15$ °C (see Fig. 4).

### 3.1.1. Average biomass and time-series in the trophic web

Climate change scenarios predict an increase of average temperature with consequent impacts on biogeochemical processes. The increase of average temperature has been associated with a reduction of carbon in the system biomass [23], in particular for herbivores [87], with cascading effects on the whole ecosystem. On the contrary, in the present analysis we perturb the variance of the temperature while keeping the average value the same as the deterministic case (see Eq. (10)).

The mean value concentrations of the populations, shown in Fig. 5, appear to depend on both $\tau$ and $D$, whose values determine the steady value of the temperature's standard deviation (amplitude of fluctuations) as shown in Fig. 4. It is possible to notice that the values reached by $\sigma_T(t)$, within a time window of ten years, are high (most of them in the top right corner of the map). Further, the mean value concentrations of the nine populations are not equally influenced by $\sigma_T$ (random fluctuations of temperature). The concentrations of the most abundant species in the stochastic approach (B1, P3, Z5, Z6, Z4) deviate from the deterministic value (see bottom left corner of the maps in Fig. 5) approximately of -13% for bacteria (B1) and -32% for omnivorous mesozooplankton (Z4), approaching the region with $\sigma_T > 10°C$ (see top right corner of the maps in Fig. 5). Larger effects ($> 100\%$) are found for P1, P2, P4, Z3.

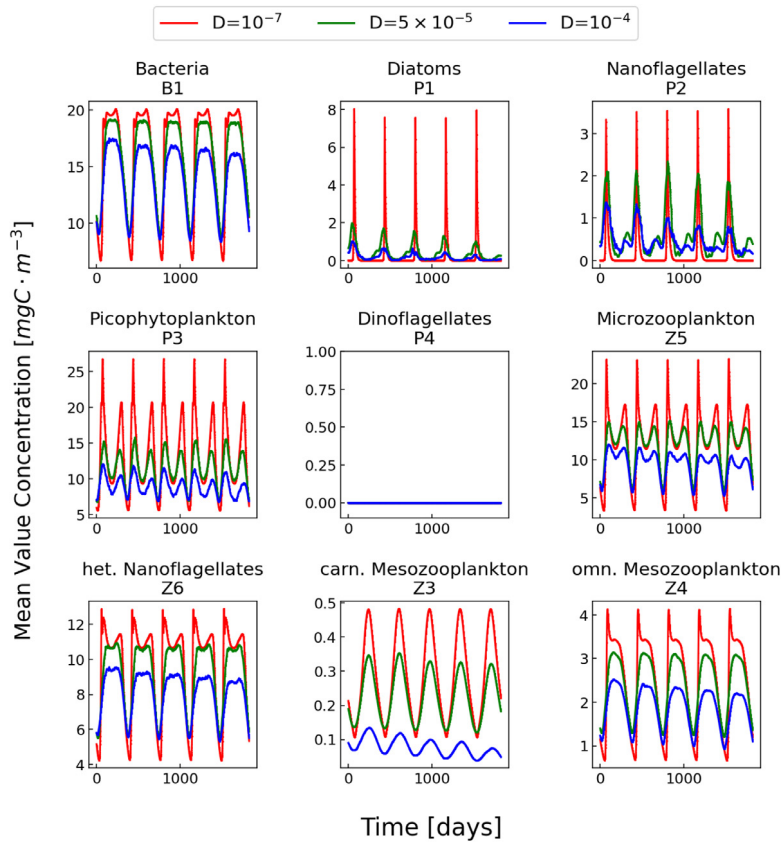

**Fig. 7.** Results from the stochastic 0-D BFM. Each panel shows the time evolution of the mean value (the average is taken over the 1000 realizations) of carbon intracellular content ($mgC/m^3$) of BFM plankton functional types. The time series shown are five-year long with daily frequency. Five-year spin-up is applied. The red, green, and blue curves are related to three different values of noise intensity, $D = 10^{-7}$ °$C^2/s$, $D = 5 \cdot 10^{-5}$ °$C^2/s$, $D = 10^{-4}$ °$C^2/s$, respectively, for $\tau = 45$ *days*.

In a linear model (*L*), if one considers a generic variable *x*, the stochastic perturbations in temperature, governed by Eq. (9), do not affect its average value, as it can be seen from the following equation:

$$\frac{d<x>}{dt} = <L(x,\xi)> = L(<x>,<\xi>) \tag{16}$$

since $<\xi> = 0$, while an effect is observed only on the variance of the biogeochemical state. On the contrary, the presence of nonlinearities causes a change in the average values of state variables. Our analysis indeed (see Fig. 5) indicates that the nonlinearity affects mostly the outcompeted primary producers (P1, P2, P4) and the top predator (Z3).

Maps of standard deviation (image not shown) exhibit a pattern qualitatively similar to that of temperature fluctuations, shown in Fig. 4.

Dinoflagellates are always extinct except for $\tau = 365\ days$ and $D \geq 7.5 \times 10^{-6}\ °C^2/s$, which indicates the constructive role of random fluctuations, whose presence can contribute to preserve some populations excluded in the deterministic regime. From the point of view of real ecosystems this could explain the appearance or disappearance of some planktonic groups and the enhancement of diversity driven by fluctuations [88].

Nonmonotonic behaviour emerges both for bacteria and microzooplankton as a function of the noise parameters, a phenomenon typical of noise-affected nonlinear systems. In particular, bacteria exhibits a nonmonotonic behaviour with respect to $\tau$ for $D = 10^{-4}\ °C^2/s$, whereas microzooplankton manifests nonmonotonicity with respect to *D* for $\tau = 365\ days$.

In order to appreciate better this point and to find out other possible nonmonotonic behaviours, in Fig. 6 different curves for the most relevant values of $\tau$ against the parameter *D* are shown. Besides the nonmonotonicity for $\tau = 365\ days$, which characterizes the microzooplankton and, to a lesser extent, bacteria and omnivorous mesozooplankton, other two nonmonotonic behaviours are present: diatoms and nanoflagellates indeed exhibit an evident maximum for $\tau$ varying in the interval [45, 90] *days*.

In the light of the above observations, in Figs. 7 and 8 we show the time series of the nine populations in the last five years of the simulation time interval for the two values of $\tau$ for which non monotonic behaviours are observed, i.e. $\tau = 45\ days$ and $\tau = 365\ days$.

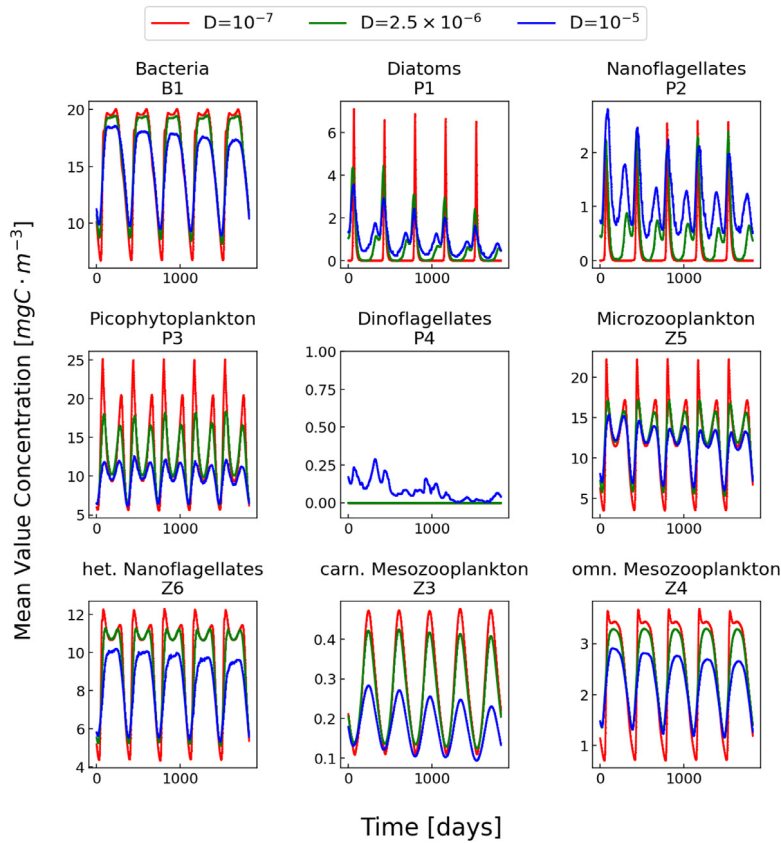

**Fig. 8.** Results from the stochastic 0-D BFM. Each panel shows the time evolution of the mean value (the average is taken over the 1000 realizations) of carbon intracellular content ($mgC/m^3$) of BFM plankton functional types. The time series shown are five-years long with daily frequency. Five-year spin-up is applied. The red, green, and blue curves are related to three different values of noise intensity, $D = 10^{-7}\ °C^2/s$, $D = 2.5 \cdot 10^{-6}\ °C^2/s$, $D = 10^{-5}\ °C^2/s$, respectively, for $\tau = 365\ days$.

In the first case ($\tau = 45$ *days*, Fig. 7) we focus the analysis on diatoms and nanoflagellates, since these populations exhibit a clear nonmonotonic behaviour, that can not be appreciated by the peaks present in the time series, whose magnitude decreases with increasing noise intensity. Rather, the presence of a nonmonotonic behaviour is more evident by looking at the mean value of the concentrations. Indeed, the red curve shows the highest peaks and the lowest mean value, being different from zero only in a very limited time-window. The green curve, instead, corresponding to the noise value for which the black curves in Fig. 6 have a maximum, presents the highest mean value.

In the second case ($\tau = 365$ *days*, Fig. 8) we find an analogous trend for the microzooplankton group. It is worth emphasizing that this value of $\tau$, for a suitable noise intensity, i.e. $D = 10^{-5}$ $°C^2/s$, corresponds to the most favourable condition for dinoflagellates. Moreover, it is clear that the blue curve, corresponding to $D = 10^{-5}$ $°C^2/s$, is the unique characterized by nonvanishing values. Moreover, we point out that no periodicity is present in the time evolution of dinoflagellates. This circumstance highlights that the dynamics of dinoflagellates is mainly driven by the random fluctuations present in the ecosystem, suggesting that dinoflagellates, differently from all other populations, survive only when noise is so intense as to push away significatively the ecosystem from attractors responsible for the exclusion of this group. Moreover, we note that the correlation time plays a crucial role in the ecosystem dynamics, in particular determining significantly different behaviours in some species such as dinoflagellates which, for the same value of $\sigma_T$ (temperature fluctuations), are clearly influenced as the value of $\tau$ varies. This fact suggests that the nonlinearity of the model comes more strongly into play when the threshold value, $\tau_{thr} = 365$ *days*, is reached.

The dynamics of the other populations, instead, even in the presence of high-intensity noise, is mainly driven by periodic oscillations, due to the seasonal forcing, present in the model, which reproduces the temperature and irradiance oscillations of real ecosystems. However, in some populations (Bacteria, Nanoflagellates, and omn. Zooplankton) the noise strongly affects the periodic regime, reducing the amplitude of oscillations and making more regular the oscillations themselves.

The noise, thus, contributes to regulate the dynamics, pushing the system towards stationary states characterized by smaller oscillations. The low noise intensity curves (red lines in Figs. 7 and 8), in fact, which are quite superimposable with the deterministic curves, exhibit high and tight peaks, indicating that the biological populations should undergo an intense change in a short time range. The stochastic curves with a sufficiently high level of noise (green and blue lines), instead, show a more regular time behaviour.

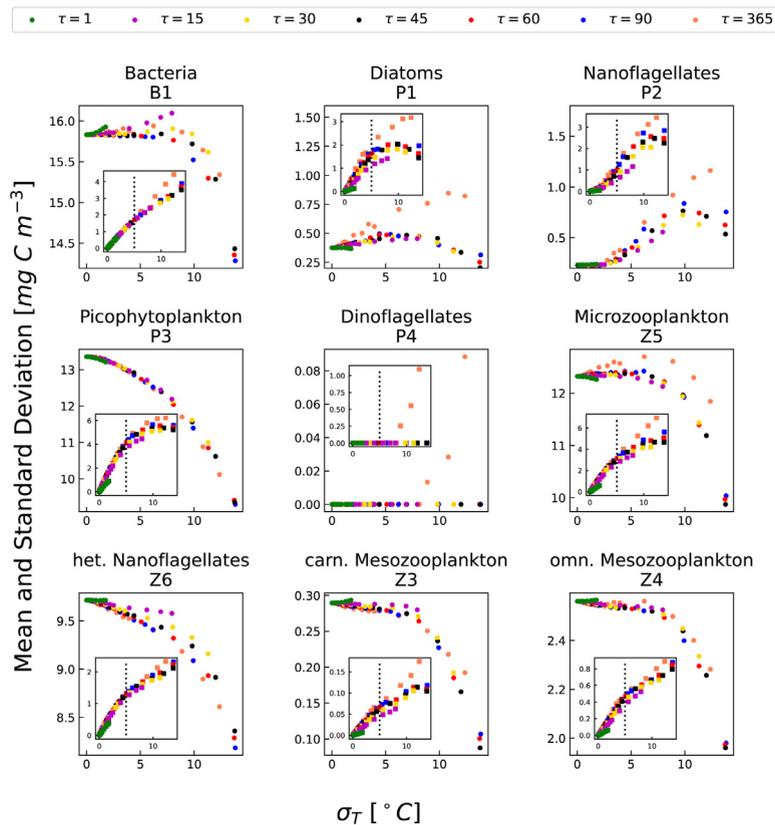

**Fig. 9.** Results from the stochastic 0-D BFM. Each panel shows the time evolution of the mean value (the average is taken over 1000 realizations) of carbon intracellular content ($mgC/m^3$) of BFM plankton functional types. $\sigma_T$ are reported in x-axis. The insets show the corresponding standard deviations, with $\sigma_T = 5°C$ being indicated by the vertical dashed black line. Different colors correspond to different values of the correlation time $\tau$.

The noise, besides making the peaks lower, causes a drastic change in the shape of the curves. In Figs. 7 and 8 one can see indeed that diatoms and nanoflagellates undergo a change from a single-peaked-per-year curve to a multi-peaked-per-year curve, whereas omnivorous mesozooplankton undergoes the opposite change. This effect can be related to the existence of, at least, two attractors in the parameter space. This change of shape in the curves can be interpreted as a possible 'transition' of the system from an attraction basin to another one. The noise can make the system explore different attraction basins, while causing transitions between different dynamical regimes. The dependence of the system on $\sigma_T$, Fig. 9, shows remarkably different behaviours for different values of $\tau$. Picophytoplankton mean values are uniquely dependent on $\sigma_T$, while standard deviation, in particular for diatoms, flagellates and dinoflagellates, strongly depends on the specific correlations times $\tau$, which shows how the interplay between nonlinearity and self-correlation time of temperature fluctuations affects the ecosystem dynamics. The standard deviations monotonically increase with $\sigma_T$. Very large coefficients of variation (the ratio of standard deviation to average) are found for diatoms and flagellates even with fluctuations of temperature below 5 °C (data not shown). Noise is, therefore, highly effective in dispersing the trajectories of species that are seasonally outcompeted in the deterministic BFM.

*3.1.2. Fourier analysis*

In this paragraph we calculate the power spectral densities (PSDs) of the population time series. A PSD consists in the square module of the Fourier transform of a signal. In our case, the signals are the concentration time series of the nine populations. Specifically, we calculate, for each stochastic realization, the square module of the Fourier transform of the time series previously obtained, then we get the mean PSDs by averaging over the 1000 realizations. This analysis gives us information about which frequencies characterize the ecosystem dynamics and how much these frequencies "weight" depending on the noise intensity.

In Fig. 10 the PSDs of the nine populations are shown for $\tau = 45$ *days* and three different values of the noise intensity, $D = 10^{-7}$ °$C^2/s$ (red curve), $5 \times 10^{-5}$ °$C^2/s$ (green curve) and $10^{-4}$ (blue curve), for which some population time series have shown a nonmonotonic behaviour. The deterministic curves ($D = 0$ °$C^2/s$) are not shown since they overlap with those obtained for the smallest value of noise intensity (see red curve).

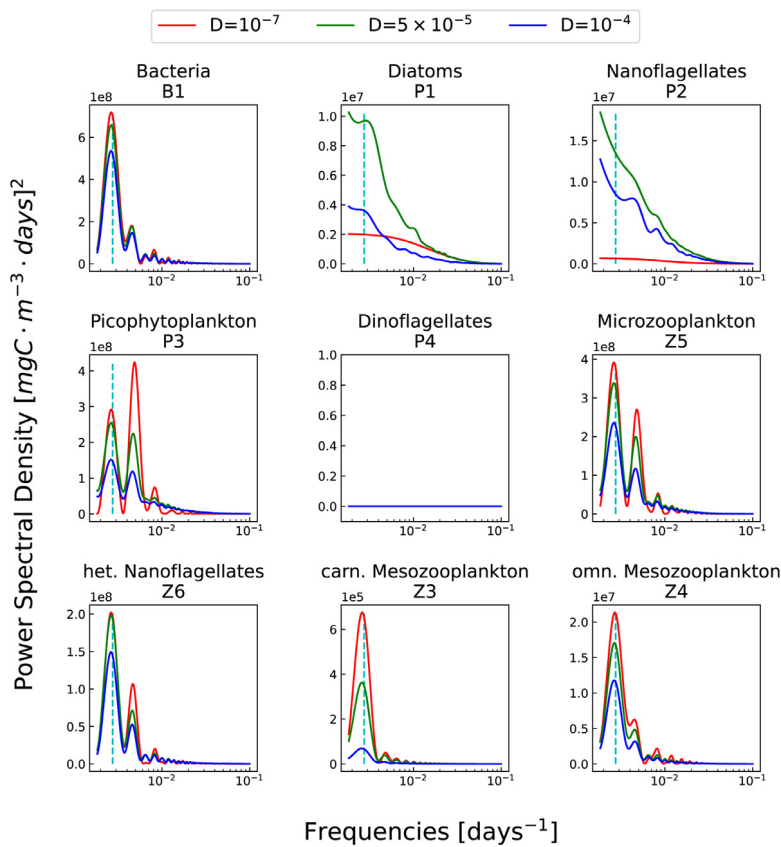

**Fig. 10.** Results from the stochastic 0-D BFM. The nine panels show the mean (the average is taken over 1000 realizations) power spectral density of the carbon intracellular content signal (Fig. 7) for each BFM plankton functional type. The frequency range considered is $[10^{-1}, 1.8 \cdot 10^{-3}]$ with a step of 1 $day^{-1}$; the extremes correspond to the frequencies related to 10 days and 1.5 years. The red, green, and blue curves are related to three different values of noise intensity, $D = 10^{-7}$ °$C^2/s$, $D = 5 \cdot 10^{-5}$ °$C^2/s$, $D = 10^{-4}$ °$C^2/s$, respectively, for $\tau = 45$ *days*. The vertical, cyan, dashed line indicates the frequency corresponding to one-year period.

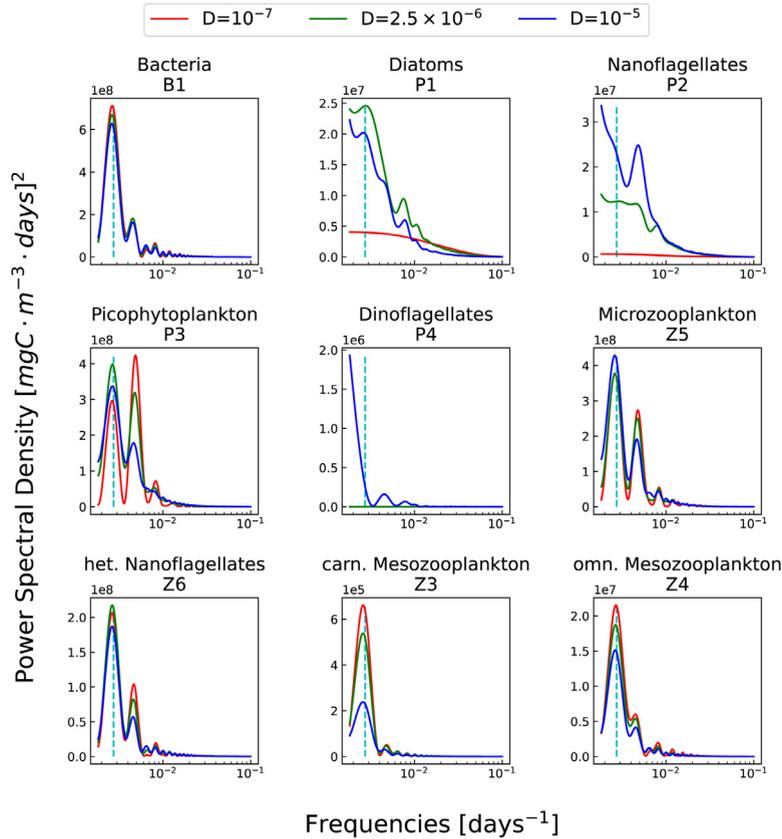

**Fig. 11.** Results from the stochastic 0-D BFM. The nine panels show the mean (the average is taken over 1000 realizations) power spectral density of the carbon intracellular content signal (Fig. 8) for each BFM plankton functional type. The frequency range considered is $[10^{-1}, 1.8 \cdot 10^{-3}]$ with a step of 1 $day^{-1}$; the extremes correspond to the frequencies related to 10 days and 1.5 years. The red, green, and blue curves are related to three different values of noise intensity, $D = 10^{-7}$ $°C^2/s$, $D = 5 \cdot 10^{-5}$ $°C^2/s$, $D = 10^{-4}$ $°C^2/s$, respectively, for $\tau = 365$ *days*. The vertical, cyan, dashed line indicates the frequency corresponding to one-year period.

In order to analyse the spectral content of the signals (time series of the species abundances), we focus on seasonal variability, restricting our analysis to a time window of one year and half, and obtaining the power spectra shown in Fig. 10 and Fig. 11. We see that almost all species present peaks at frequencies corresponding to periods less or equal to one year. Diatoms and nanoflagellates, instead, do not exhibit the secondary peaks. Rather, their spectra resemble the characteristic PSD of a Gaussian signal.

Diatoms and nanoflagellates, besides showing a different PSD shape compared to other populations, present also a different response as noise intensity varies. In particular PSDs of both diatoms and nanoflagellates present a nonmonotonic behaviour for increasing values of the noise intensity. This can be traced back to the nonmonotonicity previously discussed and observed in Fig. 6.

For both correlation times considered ($\tau = 45, 365$ *days*) no resonance between seasonal driving and noise source is observed, in contrast to what is observed in [35]. Furthermore, we note that the power spectra of the time series highlight the correlation between preys and predators, evidencing a "fingerprint" of the trophic web. It is possible to convince oneself of this by looking at carnivorous and omnivorous mesozooplankton, the two predator populations located at the top of the food chain. In particular, in the spectrum of omnivorous mesozooplankton several appreciable secondary peaks are present for frequencies higher than $\nu = \frac{1}{180} days^{-1}$. Omnivorous mesozooplankton indeed feeds on groups, like diatoms and nanoflagellates, whose spectrum shows a significant power absorption in the frequency range analyzed. On the contrary, the power spectrum of carnivorous mesozooplankton exhibits only one peak since this population feeds almost exclusively on omnivorous mesozooplankton and dinoflagellates, whose concentrations however mostly vanish in this ecosystem.

## 4. Concluding remarks

We have developed a Stochastic Biogeochemical Flux Model, by introducing additive noise terms in the deterministic BFM, in view of taking into account both the contribution of random fluctuations coming from the environment and eventual uncertainties such as intrinsic variabilities [43].

We wish to point out that additive noise sources could be used to better model random fluctuations of: i) parameters which regulate the kinetics of a given chemical process; ii) the rate of transition between two different phases caused by a certain contaminant (for example from Hg dissolved to particulate Hg); iii) the rate at which a biological population absorbs a contaminant; iv) the so called recycling rate, that is, the rate at which the amount of a contaminant is reintroduced into the ecosystem since it is absorbed by a biological population.

The SBFM introduced and used in this work presents additive noise terms to model random fluctuations of the temperature. In particular, we have considered an Ornstein-Uhlenbeck process in order to mimic memory effects, which contribute to limit the amplitude of temperature fluctuations, preventing them from becoming excessively large, which would be in contrast with the dynamics of a real ecosystem. Moreover, as a first attempt, a monotonic dependence of the temperature-regulating factor has been taken into account.

We recall that it is extremely interesting to study the effects of noise in marine ecosystems, which are a particular example of complex system, because of their intrinsic nonlinearity and their interaction with random fluctuations coming from the environment. The interplay between the nonlinearity and noisy perturbations indeed makes the system dynamics unpredictable in a non-trivial way, giving rise to counterintuitive phenomena such as stochastic resonance [46–49], noise enhanced stability [50,51] and noise delayed extinction [34,52], typical of nonlinear and noisy systems. A practical remarkable example consists, as previously noted, in marine ecosystems since their dynamics is governed by both deterministic forcings (daily and seasonal cycles) [26,54,55] and random fluctuations of environmental variables [30,34,42,58] such as the temperature [59,60]. The first counterintuitive noise-induced effect on the population dynamics described by the SBFM model is the nonmonotonic behaviour of the biomass concentration with respect to both the intensity $D$ and the correlation time $\tau$ of the noise source. More precisely the study involves the mean value of the biomass concentrations, which is calculated by averaging over the stochastic realizations, whose number were set to 1000. Moreover, we calculate the average over the whole simulation time. This allowed to highlight for bacteria and microzooplankton a nonmonotonic behaviour of their average biomass concentrations: i) as a function of $\tau$ for $D = 10^{-4}$ $°C^2/s$, a noise intensity corresponding to a realistic amplitude of temperature fluctuations; ii) as a function of $D$ for $\tau = 365$ *days*, a quite long correlation time, which coincides with the seasonal periodicity. These behaviours are observed also for other values of $D$ and $\tau$, in particular in diatoms and nanoflagellates, as a function of $D$ for three different values of $\tau$. This nonmonotonicity can be better appreciated through the power spectral analysis.

Another remarkable aspect emerging by increasing the noise intensity is the damping of oscillations. This effect can be interpreted as a regulating function of the noise. That is, the noise makes the oscillations smoother with a consequent more balanced system dynamics, which allows to get more scenarios in view of describing real ecosystems through model calibration *versus* dataset.

Such a regulating function can be found also in the corresponding PSDs. The spectral analysis brought to light a feature which characterizes the complex trophic networks [89,90]: some groups (B1, P3) of the lower trophic level present pronounced secondary peaks, meaning that their dynamics is characterized by different frequencies. On the contrary, in the higher trophic levels the species (Z3 and Z4) are characterized by a more regular time evolution. In this case, the time series closely resemble a sinusoid as confirmed by the power spectra, where just one frequency, corresponding to the highest peak, is present, with a secondary peak representing the doubled main frequency, that is the first higher harmonic.

It is also worth to remark a rather surprising effect stemming from the randomly varying temperature. For $\tau = 365$ *days* and $D = 10^{-5}$ $°C^2/s$ we observe a noise delayed extinction for the Dinoflagellate group. They are always extinct both in the deterministic case and in the most part of stochastic scenarios. Only when noise is strong enough and the correlation frequency of the forcing is sufficiently high, dinoflagellates succeed in surviving for a longer time, precisely during the whole observation time. This indicates that such a planktonic group takes advantage from environmental random fluctuations. The dynamics of dinoflagellates indeed is clearly noise driven, as it shows no evident seasonal forcing-induced time behaviour.

As a general result, we wish to note that this work showed how the interplay between nonlinearity and environmental random fluctuations allows a deeper comprehension of the ecosystem dynamics. Possible outlooks stemming from the present work can be: i) exploring the effects and the impact on the population dynamics when a non-monotonic temperature-regulating factor is considered; ii) studying the observed phenomenology, when other physical or chemical variables, deeply influencing the system dynamics such as the light, are modelled as stochastic processes. We conclude noting that it could be a rich source of information analyzing the system response to a multiplicative noise which directly affects the population concentrations.

**Declaration of Competing Interest**

The authors declare that they have no known competing financial interests or personal relationships that could have appeared to influence the work reported in this paper.

**CRediT authorship contribution statement**

**Paolo Lazzari:** Conceptualization, Methodology, Software, Formal analysis, Investigation, Resources, Writing – original draft, Writing – review & editing. **Roberto Grimaudo:** Methodology, Software, Formal analysis, Investigation, Resources,

Writing – original draft, Writing – review & editing. **Cosimo Solidoro:** Conceptualization, Methodology, Investigation, Resources, Writing – original draft, Writing – review & editing, Supervision, Project administration, Funding acquisition. **Davide Valenti:** Conceptualization, Methodology, Investigation, Resources, Software, Writing – original draft, Writing – review & editing, Supervision, Project administration, Funding acquisition.

## Acknowledgments

The work was carried out within the PRIN Project PRJ-0232 - Impact of Climate Change on the biogeochemistry of Contaminants in the Mediterranean sea (ICCC). The authors acknowledge the financial support of the Ministry of University and Research of Italian Government. The authors are thankful to Duilio De Santis for useful discussions and suggestions.